\documentclass{INTERSPEECH2023}
\usepackage{algorithm}
\usepackage{algpseudocode}
\usepackage{multirow}
\usepackage{array}
\usepackage{url}
\usepackage{cite}
\usepackage{bm}
\usepackage{color}

\interspeechcameraready

\title{Model Compression for DNN-based Speaker Verification Using Weight Quantization}
\name{Jingyu Li$^1$, Wei Liu$^1$, Zhaoyang Zhang$^1$, Jiong Wang$^2$, Tan Lee$^1$}
\address{
  $^1$Department of Electronic Engineering, The Chinese University of Hong Kong, Hong Kong\\
  $^2$School of Science and Engineering, The Chinese University of Hong Kong, Shenzhen, China}
\email{\{lijingyu0125, louislau\_1129, zhaoyangzhang\}@link.cuhk.edu.hk, jiongwang@link.cuhk.edu.cn, tanlee@ee.cuhk.edu.hk}

\begin{document}

\maketitle
 
\begin{abstract}
DNN-based speaker verification (SV) models demonstrate significant performance at relatively high computation costs. Model compression can be applied to reduce the model size for lower resource consumption. The present study exploits weight quantization to compress two widely-used SV models, namely ECAPA-TDNN and ResNet. Experimental results on VoxCeleb show that weight quantization is effective for compressing SV models. The model size can be reduced multiple times without noticeable degradation in performance. Compression of ResNet shows more robust results than ECAPA-TDNN with lower-bitwidth quantization. Analysis of the layer weights suggests that the smooth weight distribution of ResNet may be related to its better robustness. The generalization ability of the quantized model is validated via a language-mismatched SV task. Furthermore, analysis by information probing reveals that the quantized models can retain most of the speaker-relevant knowledge learned by the original models.

\end{abstract}

\section{Introduction}
\label{sec:intro}

Speaker verification (SV) is a biometric authentication process verifying whether a spoken utterance is from the claimed speaker\cite{campbell1997speaker,wu2015spoofing}. Deep neural network (DNN) models have shown great success in SV\cite{snyder2018x,Nagrani17,desplanques2020ecapa,thienpondt21_interspeech} and outperformed conventional methods, e.g., I-vector\cite{dehak2010front}. Despite significant performance attained with these DNN models, it comes at the cost of higher computation complexity.
SV systems are expected to be run locally on devices without the Internet\cite{DBLP:conf/interspeech/KimCK21}. In this way, users' private data are kept safe without being exposed outside the device. The data transmission delay can also be avoided.  However, the demanding computation cost of current DNN-based SV systems makes them not handy to deploy on resource-constrained devices.

To reduce resource consumption, model compression is needed. The performance achieved by the compressed model should remain comparable to the original model. Various compression methods, such as weight quantization\cite{choi2018pact,DBLP:conf/iclr/LiDW20,DBLP:conf/icml/ZhangSG0L21}, pruning\cite{han2015deep,DBLP:conf/iclr/0022KDSG17}, and knowledge distillation\cite{DBLP:conf/iclr/PolinoPA18}, have been studied on different tasks\cite{salvi2021model,cheng2018model}. The present study focuses on weight quantization for DNN-based speaker embedding extraction models. The default data bitwidth used in existing deep learning frameworks is 32, e.g., PyTorch\cite{paszke2019pytorch} and TensorFlow\cite{tensorflow2015-whitepaper}. The core operation in weight quantization is mapping full-bitwidth model weights onto a lower-bitwidth representation, which reduces the model size and enables faster processing. 


To the best of our knowledge, weight quantization has not been investigated on state-of-the-art DNN SV models. ECAPA-TDNN\cite{desplanques2020ecapa} and ResNet\cite{he2016deep,Nagrani17} are two widely-used well-performing models for speaker embedding extraction. Many SV models have been developed based on them\cite{DBLP:conf/icassp/LiuDLL22,thienpondt21_interspeech,li2020text}. In this study, we evaluate the effect of weight quantization on these two models. The experiments are carried out on the VoxCeleb datasets\cite{Nagrani17,Chung18b,Nagrani19}, with two common learnable quantization methods: uniform quantization\cite{cai2017deep} and Powers-of-Two quantization\cite{DBLP:conf/iclr/ZhouYGXC17}. Our preliminary experiments show that weight quantization is effective for SV model compression. With multiple times smaller sizes than the original models, the performance of quantized ECAPA and ResNet declines slightly.
ResNet shows  more robust results than ECAPA, especially in 4-bit quantization. To investigate the cause of such a difference, we analyze the layer-wise weight distribution of these two models. The highly zero-concentrated and sharp weight distribution of ECAPA is observed, which may explain its less robustness than ResNet using lower-bitwidth quantization. 
As reducing the model size, the learning ability of the quantized model is decreased compared to the original counterpart. The poorer learning ability may affect the models' generalization capability and extracted knowledge. We evaluate the generalization of quantized models using a language-mismatched SV evaluation, which is performed on CN-Celeb\cite{fan2020cn} without model fine-tuning. Inspired by \cite{raj2019probing}, four information probe tasks are applied to the extracted speaker embeddings to investigate the model's learned knowledge before and after quantization.

The paper is organized as follows: Section~\ref{sec:quantization} introduces the details of the quantization methods. The experimental setting and results are given in Sections~\ref{sec:exp_setting} and ~\ref{sec:exp_result}, respectively. Finally, we give a brief conclusion in Section~\ref{sec:conclusion}.

\section{Quantization}
\label{sec:quantization}

A trainable weight in a layer is denoted as $\mathcal{W} \in \mathbb{R}^N$. In a 2D convolutional layer (CNN), $N=k\times k \times c_1\times c_2$. In a fully-connected layer (FC), $N=c_1\times c_2$. $\mathcal{W}$ takes continuous values. The quantization of $\mathcal{W}$ is expressed as
\begin{equation}
  \mathcal{\widehat{W}}=\Gamma(\lfloor \mathcal{W},\alpha \rceil)_{\bm{q}(\alpha, b)}.
  \label{eq:quant}
\end{equation}
$\mathcal{W}$ is first normalized by its mean and standard deviation
so that it can be represented in a symmetric range. They are clipped into a range of $[-\alpha, \alpha]$ by a function denoted as $\lfloor \cdot,\alpha \rceil$, where $\alpha\ (\geq 0)$ is a learnable parameter. $\Gamma(\cdot)_{\bm{q}}$ represents a projection function, mapping continuous values into a set of discrete values $\bm{q} = [\pm q_1, \pm q_2,..., \pm q_{\frac{n}{2}}]$. 
$n=2^b$ is the number of quantization levels. $b$ is the bitwidth and is usually pre-determined.
The quantization level, i.e., the values of $[q_1, q_2...]$, is another critical factor determining the performance of the quantized models. Uniform and Powers-of-Two quantization are widely used\cite{cai2017deep,gong2019differentiable,miyashita2016convolutional,DBLP:conf/iclr/ZhouYGXC17,gholami2021survey} for constructing the quantization level. They are friendly to both software and hardware implementations. 

\begin{figure}[t]
  \centering
  \includegraphics[width=\linewidth]{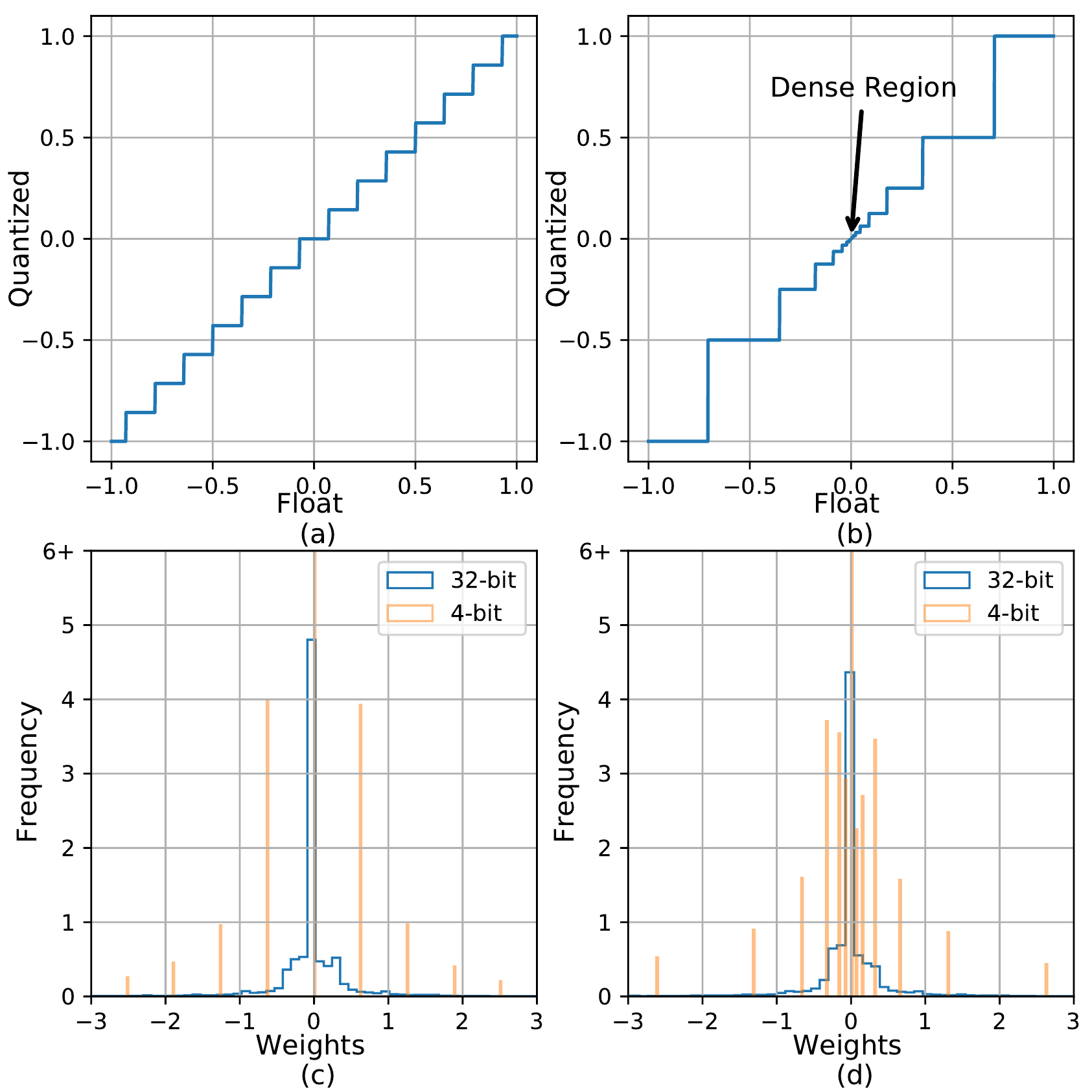}
  \vspace{-4mm}
  \caption{The quantized values of (a) uniform and (b) PoT quantization versus original float32 values. The weights distribution of 32-bit weights and 4-bit quantized by (c) uniform, (d) PoT.}
  \label{fig:quant}
  \vspace{-3mm}
\end{figure}

\subsection{Uniform quantization}
  \vspace{-1mm}
\label{ssec:uni_Q}
The quantization levels of uniform quantization are represented as
\vspace{-1mm}
\begin{equation}
  \bm{q}(\alpha, b) = [0,\frac{\pm 1}{2^{b-1}-1},\frac{\pm 2}{2^{b-1}-1},...,\pm 1] \times \alpha.
  \label{eq:uni_quant}
\end{equation}
Uniform quantization generates the discrete values evenly over the interval $[-1, 1]$, as illustrated in Fig.~\ref{fig:quant}(a). The length of each quantization step is constant, i.e., $\frac{1}{2^{b-1}-1}$. A smaller step length gives denser discrete levels to distinguish different values.
However, the distribution of weights is usually non-uniform\cite{han2015deep}. For instance, a layer's weight distribution from a trained model is plotted by the blue curve in Fig.~\ref{fig:quant}(c). There are more weights concentrated around the center and fewer weights around the boundaries. The orange line in Fig.~\ref{fig:quant}(c) shows the distribution of uniform-quantized weights. The minor numerical differences between weights are erased if they lie in the same quantization level. Thus a large proportion of the weights around the center will be assigned to an identical value, which may dramatically decrease the diversity of weights.

\subsection{Powers-of-Two quantization}
  \vspace{-1mm}
\label{ssec:pot_Q}
 The quantization levels of Powers-of-Two (PoT) quantization are represented as
\vspace{-1mm}
\begin{equation}
  \bm{q}(\alpha, b) = [0, \pm 2^{-2^{b-1}+2}, ..., \pm 2^{-1}, \pm 1] \times \alpha.
  \label{eq:pot_quant}
\end{equation}
PoT provides non-identical discrete levels, unlike uniform quantization, as shown in Fig.~\ref{fig:quant}(b). The quantization levels around the center are denser. The denser levels provide a more refined quantization description of weight parameters, as shown by the orange lines in Fig.~\ref{fig:quant}(d). Sparser quantization levels are proposed near the boundary. If only a few weights lie around the boundary, as in Fig.~\ref{fig:quant}(d), discretizing the weights may not significantly affect the quantization error  ($||\mathcal{W}-\mathcal{\widehat{W}}||^2$). 

\begin{table*}[t]
  \caption{Performances of the models. Vox.O, Vox.E, Vox.H are short for the Original, Easy and Hard test sets of Vox.1}
  \vspace{-1mm}
  \label{tab:results}
  \setlength\tabcolsep{7pt}
  \centering
  \begin{tabular}{c c c c c c c c c}
    \toprule
    \multirow{2}{*}{\textbf{Model}} & \textbf{Parameters}  & \textbf{MACs}  & \multirow{2}{*}{\textbf{Compression}}  & \textbf{Bitwidth} & \textbf{Size} & \multicolumn{3}{c}{\textbf{EER(\%)}} \\
    
    \cline{7-9}
      & \textbf{(M)} & \textbf{(Giga)} & & \textbf{(bits)} & \textbf{(MBytes)} & \textbf{Vox.O} & \textbf{Vox.E} & \textbf{Vox.H}  \\
    \midrule
    \midrule
    \multirow{5}{*}{ECAPA-TDNN-512} & \multirow{5}{*}{5.95} & \multirow{5}{*}{0.96} & -   & 32 & 23.87 & 1.07 & 1.28 & 2.43 \\
    \cline{4-9}
                                 &   & & Uni & \multirow{2}{*}{8}   & \multirow{2}{*}{6.09}  & 1.20 & 1.46 & 2.75 \\
                                 &   & & PoT &                      &                        & 1.44 & 1.69 & 3.11 \\
    \cline{4-9}                             
                                 &   & & Uni & \multirow{2}{*}{4}   & \multirow{2}{*}{3.13}  & 1.56 & 1.78 & 3.29 \\
                                 &   & & PoT &                      &                        & 1.65 & 1.86 & 3.41 \\
    \midrule
    \multirow{3}{*}{ECAPA-TDNN-1024} & \multirow{3}{*}{14.38} & \multirow{3}{*}{2.56} & - & 32  & 57.61 & 0.92 & 1.15 & 2.28 \\
    \cline{4-9}
                                 &    &    & Uni & \multirow{2}{*}{8}   & \multirow{2}{*}{14.6} & 1.26 & 1.45 & 2.73 \\
                                 &    &    & PoT &                      &  & 1.41 & 1.59 & 2.99 \\
    \midrule
    \multirow{5}{*}{ResNet34} & \multirow{5}{*}{6.9}  & \multirow{5}{*}{3.67} & - & 32  & 27.63 & 1.10 & 1.15 & 2.08 \\
    \cline{4-9}
                            &  & &  Uni & \multirow{2}{*}{8}    & \multirow{2}{*}{6.96} & 1.05 & 1.11 & 2.02 \\
                            &  & &  PoT &                       &  & 1.28 & 1.34 & 2.41 \\
    \cline{4-9}
                            &  & &  Uni & \multirow{2}{*}{4}    & \multirow{2}{*}{3.52} & 1.22 & 1.29 & 2.32 \\
                            &  & &  PoT &                       &  & 1.35 & 1.40 & 2.52 \\
    \midrule
    \multirow{3}{*}{ECAPA-TDNN-256} & \multirow{3}{*}{3.11} & \multirow{3}{*}{0.43} & - & \multirow{3}{*}{32} & \multirow{3}{*}{12.5} & 1.43 & 1.59 & 2.87 \\
                                 &   &                       & distill(512)  &         &  & 1.29 & 1.53 & 2.88 \\
                                &    &                       & distill(1024) &         &  & 1.30 & 1.53 & 2.85 \\
    \midrule
    \multirow{2}{*}{ResNet18} & \multirow{2}{*}{4.37} & \multirow{2}{*}{1.78} & - & \multirow{2}{*}{32}  & \multirow{2}{*}{17.51} & 1.45 & 1.44 & 2.53 \\
                            &  &    & distill             &                      &  & 1.33 & 1.39 & 2.48 \\
    \bottomrule
  \end{tabular}
  \vspace{-2mm}
\end{table*}

\subsection{Parameter update}
\vspace{-2mm}
\label{ssec:quant_train}
The quantization error ($||\mathcal{W}-\mathcal{\widehat{W}}||^2$) depends heavily on the value of $\alpha$. A large $\alpha$ covers a wide range of weight values, and the quantization levels are distributed sparsely. A smaller value of $\alpha$ gives denser discrete levels, and more weights will be clipped to $-\alpha$ or $\alpha$, resulting in higher quantization errors. The distributions of weights in different layers are not identical, and a constant clipping value may not be optimal for quantizing $\mathcal{W}$. In this paper,  $\alpha$ is made learnable as in \cite{choi2018pact}. Each layer is assigned an $\alpha$, and the value of $\alpha$ is updated during model training by backward propagation. $\lfloor \cdot,\alpha \rceil$ and $\Gamma(\cdot)_{\bm{q}}$ are two discrete operations, where the gradients cannot be calculated directly. Following \cite{DBLP:conf/iclr/LiDW20}, the gradients for $\alpha$ and $\mathcal{W}$ are estimated by Straight-Through Estimator (STE)\cite{bengio2013estimating} as
\begin{equation}
  \frac{\partial \mathcal{L}}{\partial \alpha}=\frac{\partial \mathcal{L}}{\partial \mathcal{\widehat{W}}}
  \frac{\partial \mathcal{\widehat{W}}}{\partial \alpha},\   \frac{\partial \mathcal{\widehat{W}}}{\partial \alpha} =\begin{cases}
                                                                              sign(\mathcal{W}) & \text{if $|\mathcal{W}| > \alpha$}\\
 \frac{\mathcal{\widehat{W}}}{\alpha} - \frac{\mathcal{W}}{\alpha} & \text{if $|\mathcal{W}| \leq \alpha$}
                                                                            \end{cases},
  \label{eq:a_grad}
\end{equation}
  \vspace{-2mm}
\begin{equation}
  \frac{\partial \mathcal{L}}{\partial \mathcal{W}}=\frac{\partial \mathcal{L}}{\partial \mathcal{\widehat{W}}}
  \frac{\partial \mathcal{\widehat{W}}}{\partial \mathcal{W}},\   \frac{\partial \mathcal{\widehat{W}}}{\partial \mathcal{W}}=1.
  \label{eq:w_grad}
\end{equation}
where $\mathcal{L}$ is the training loss. In each training step, the full-precision $\mathcal{W}$ is first converted into  $\mathcal{\widehat{W}}$ with discrete values. $\mathcal{\widehat{W}}$ is used for the computational process, such as convolution or matrix multiplication. Thus the calculation of $\frac{\partial \mathcal{L}}{\partial \mathcal{\widehat{W}}}$ can be performed as the conventional gradient computation. After training, the $\mathcal{W}$ is quantized to $\mathcal{\widehat{W}}$ and stored. In the inference, 
the model with weight $\mathcal{\widehat{W}}$ is used to perform the forward process, which requires less device memory and computation cost.

\section{Experimental Set-up}
\label{sec:exp_setting}
  \vspace{-1mm}
\subsection{Datasets}
\label{ssec:datasets}
  \vspace{-1mm}
The datasets utilized in the experiments are VoxCeleb $1\&2$ (denoted as Vox.1 $\&$ Vox.2) \cite{Nagrani17,Chung18b,Nagrani19}. Speech data in VoxCeleb are mostly in English. The development set of Vox.2 consists of $5,994$ speakers and is used for model training. Vox.1 is used to evaluate the performance of models. Three test sets are constructed using data from Vox.1 and denoted as Original, Easy, and Hard sets, respectively. 

In addition, the test set of CN-Celeb\cite{fan2020cn} is used to evaluate the models' performance in another language. CN-Celeb is a large-scale Chinese dataset consisting of about $1,000$ speakers.

\subsection{Model training and evaluation}
\label{ssec:training}
Experiments are performed on ECAPA-TDNN with 512 and 1024 channels, ResNet34 with 32 channels. The training process has two stages. The model is first trained in full precision, i.e., 32-bit in PyTorch. The model is trained to predict the input speaker identity using the AAM-softmax\cite{deng2019arcface} as the classification loss.
The margin and scale of AAM-softmax are set to be $0.2$ and $30$. The Adam optimizer\cite{DBLP:journals/corr/KingmaB14} with a weight decay of $2e\textit{-}5$ is utilized. The learning rate is initialized as $0.001$ and decayed by a ratio of $0.1$ at the 20th and 32nd epoch, respectively. 
The whole training process consists of $40$ epochs. The batch size is set to $128$. A 2-second long segment is cropped from each input speech randomly. The model input is $64$-dimensional $log$ Mel-filterbanks (FBanks) transformed from the raw waveform of each segment using the hamming window. The window size is $25ms$, and the hop length equals $10ms$. Background sound addition\cite{snyder2015musan} and audio reverberation\cite{ko2017study} are utilized for data augmentation in training. 

In the second stage, the quantized model is fine-tuned from the full-precision model for 20 epochs. The learning rate is decayed at the $10th$ and $16th$ epochs. Reducing the model size decreases the learning ability. Therefore, data augmentation is not used for training the quantized models. The rest of the training settings remain the same as the first stage. 

For evaluation, each utterance is divided into 4-second long segments, with a 1-second overlap between two adjacent segments. The cosine similarity values between the test utterance segments and the enrollment segments are averaged as the verification score. Adaptive s-norm (AS-norm)\cite{matejka2017analysis} is applied to calibrate the scores. The performance is reported in terms of equal error rate (EER).

\begin{figure}[t]
  \centering
  \includegraphics[width=\linewidth]{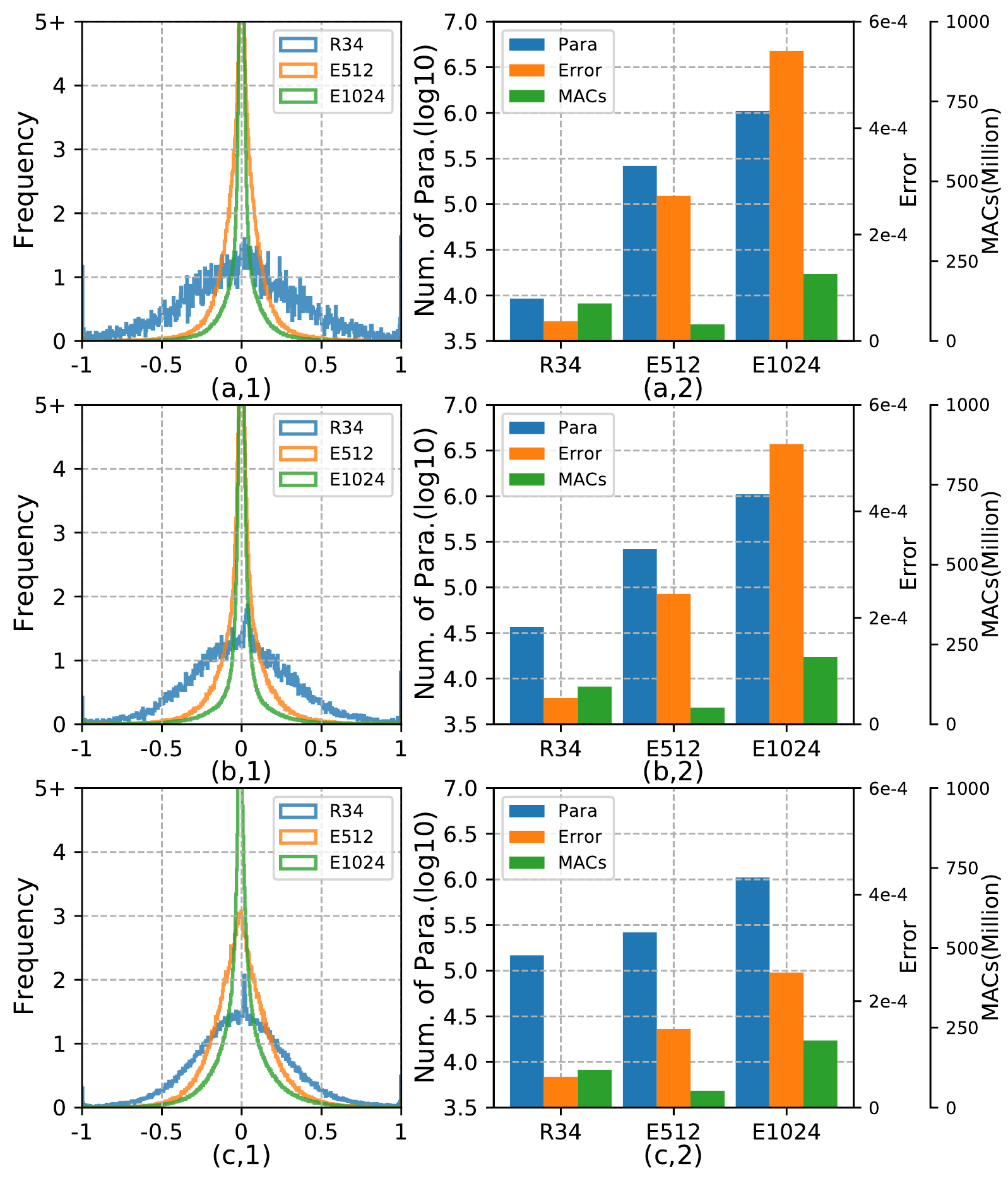}
  \vspace{-4mm}
  \caption{The weight distributions are shown in (a,1) to (c,1). R34 stands for ResNet34, E512 for ECAPA-512, and E1024 for ECAPA-1024. (a,2) to (c,2) gives the number of parameters, averaged quantization error, and MACs of layers.}
  \label{fig:layer_dis}
  \vspace{-2mm}
\end{figure}

\begin{table}[t]
  \caption{Performances of the models on CN-Celeb}
  \vspace{-1mm}
  \label{tab:CN}
  \centering
 \setlength\tabcolsep{7pt}
  \scalebox{0.99}{
  \begin{tabular}{ccc}
    \toprule
    \textbf{Model}                      & \textbf{Compression}    & \textbf{EER(\%)} \\  
    \midrule
    \midrule
    \multirow{3}{*}{ECAPA-TDNN-512}     & -                       & 17.61            \\
                                        & Uni,8-bit               & 17.50            \\
                                        & Uni,4-bit               & 16.12            \\
    \midrule
    \multirow{3}{*}{ResNet34}           & -                       & 11.85            \\
                                        & Uni,8-bit               & 11.75            \\
                                        & Uni,4-bit               & 11.93            \\
    \midrule
    ECAPA-TDNN-256                     & distill(512)            & 18.13            \\                                
    \midrule
    ResNet18                           & distill                 & 12.32            \\
    \bottomrule
  \end{tabular}}
  \vspace{-3mm}
\end{table}

\begin{figure}[t]
  \centering
  \includegraphics[width=\linewidth]{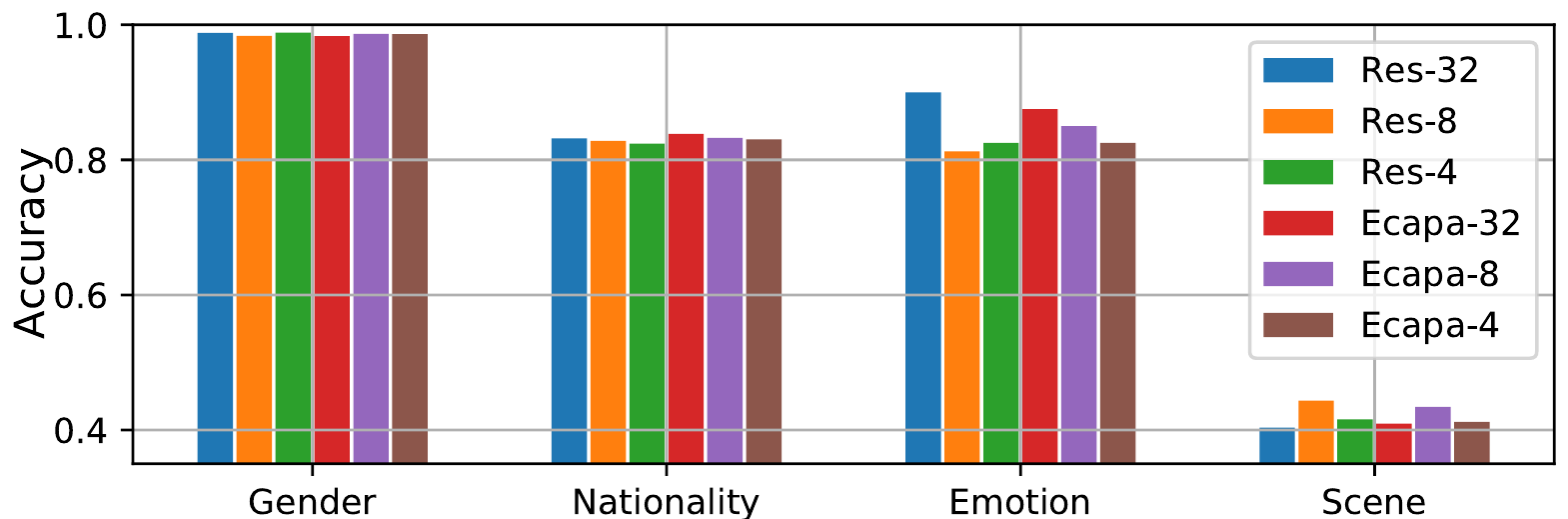}
  \vspace{-3mm}
  \caption{Accuracy of probing tasks for models with different bitwidths. Res-32 is short for ResNet34 with 32-bit, Ecapa-8 for ECAPA-512 with 8-bit, etc.}
  \label{fig:probe}
  \vspace{-3mm}
\end{figure}

\vspace{-1mm}
\section{Results and Analysis}
\label{sec:exp_result}
\vspace{-1mm}
\subsection{Learnable quantization}
\label{ssec:learnable}


The results with different models on the three test sets of Vox.1 are shown in Table~\ref{tab:results}. The computation complexity is measured in terms of the required number of multiply–accumulate operations (MACs). 8-bit quantization reduces the model size by approximately 4 times. The performance of ECAPA-TDNN-512 relatively declines by 15$\%$ on all test sets using uniform quantization. 
ECAPA-1024 outperforms ECAPA-512 with full precision but gives close results with 8-bit uniform quantization.
ResNet gives more robust results than ECAPA. The performance of ResNet34 is almost not affected with uniform quantization. PoT shows slightly poorer results than uniform on all models and test sets. The EER of PoT's output is 15$\%$ to 20$\%$ higher than the uniform's.

To evaluate the model performance under higher compression rate, 4-bit quantization is applied to quantize ECAPA-512 and ResNet34. The quantization reduces the models' size by 8 times. 
The EERs of 4-bit quantized ECAPA-512 are increased by about $50\%$ compared with the full-precision model.
The performance degradation of ResNet34 is 10$\%$ to 20$\%$, significantly milder than that of ECAPA-512. 
ResNet34 has a model size similar to ECAPA-512, but its MACs are much larger than ECAPA's, which may contribute to its minor performance degradation.

We compare the quantized models with models compressed by knowledge distillation. The distillation is performed on ECAPA-TDNN-256 and ResNet18. ECAPA-256 is constructed by using 256 convolution channels. 
As shown in Table~\ref{tab:results}, the distilled models give around $10\%$ lower EER than the model trained from scratch on all test sets. The 8-bit quantized ECAPA structure uses a half size and gives a close performance to the distilled ECAPA-256. The quantized ResNet uses a smaller size and achieves superior performance to its distilled variant.

\vspace{-1mm}
\subsection{Weight analysis}
\label{ssec:analysis}
The above results have shown that uniform quantization is more suitable than PoT on these two SV models. Here, 8-bit uniform-quantized models are utilized as representatives in weight analysis. Several layers' weight distribution and statistics information are given in Fig.~\ref{fig:layer_dis}. The last convolution layers of Block 1 to 3 of ResNet34 are given in (a)-(c), respectively. For ECAPA structures, the last convolution layers of Block 1 to 3 are shown. The weights are clipped by the corresponding $\alpha$ and scaled into [-1, 1] for better visualization. ``Error" denotes the average quantization error of the corresponding layer.

Among the three models, the weight distributions of the ECAPA networks are highly concentrated, and those of ResNet34  show smooth bell shapes. The spiky-shape distribution may lead to larger quantization error in ECAPA when uniform quantization is applied, as explained in Section~\ref{ssec:uni_Q}.
It can also be observed that the average quantization error is positively correlated with the number of parameters in a layer. Thus quantizing layers with a large number of parameters is challenging. 
The parameters are evenly distributed among layers in the same block of ResNet34, whereas the parameters concentrate densely on a block's first and last convolution layers in the ECAPA structure. The dense layers of ECAPA may explain its larger quantization performance decline compared to ResNet. This suggests a good strategy for model design is to avoid concentrating a large number of parameters on several layers.

\vspace{-2mm}
\subsection{Evaluation on CN-Celeb}
\label{ssec:generalization}
The models trained on VoxCeleb are tested on CN-Celeb without fine-tuning to evaluate their generalization ability in a different language.
The results are summarized in Table~\ref{tab:CN}. The 8-bit quantized model shows superior performance to the full-precision model on ECAPA-512, and the 4-bit model gives the lowest EER. ResNet34 outperforms ECAPA-512 with or without the quantization, indicating it has a better generalization ability. The 8-bit quantized ResNet34 also outperforms its original model. There is a slight performance drop on the 4-bit ResNet34. \cite{DBLP:conf/iclr/LiDW20} found that quantization may contribute to weight regularization, which can alleviate the overfitting problem of the full-precision model. This effect may increase the model generalization in cross-language SV. 
Performance decline is observed in the distillation models, showing that their generalization ability is worse than quantized models'.


\vspace{-2mm}
\subsection{Information probe}
\label{ssec:probe}

Information probing is conducted on the speaker embeddings extracted by the original and quantized ResNet34 and ECAPA-512. Four classification tasks are applied, i.e., gender, nationality, emotion, and scene. 
Gender and nationality are two speaker-dependent information. They are shown to be highly relevant to the speakers' variation\cite{luu2022investigating}. The gender and nationality classification tasks are trained on Vox.2 and evaluated on Vox.1. Emotions are speaker-independent. Different emotions are varied from various pitches, loudness, etc\cite{lausen2020emotion}. Previous works\cite{williams2019disentangling,pappagari2020x} have shown that speaker embeddings contain significant emotional information. SAVEE\cite{jackson2014surrey} is used for emotion classification. Scene information is expected to be speaker-irrelevant. Scene classification is performed on TAU Urban Acoustic Scenes 2020 Mobile\cite{DBLP:conf/dcase/HeittolaMV20}. A multi-layer perceptron with two FC layers and ReLU activation is utilized as the classifier for all tasks.

The results are summarized in Fig.~\ref{fig:probe}. The accuracy of gender classification is over $98\%$ for all models, showing that the quantization does not lose this information. 
Nationality and emotion are two factors related to the speaking styles of the input speech. 
The classifiers achieve accuracy higher than $80\%$ on both tasks for all models, while the accuracy is decreased slightly in the quantized models compared to the full-size models. The results indicate that the embeddings capture sufficient knowledge to distinguish different speaking styles, and a small fraction of styling information is lost in the quantization.
The scene classification accuracy is observably lower than other tasks.
However, the quantized models show higher accuracy in this task than the original models. The scene sounds may have biases for different speakers. The quantized models can rely on biased environment information for distinguishing speakers, but this effect will cause more errors in SV under noisy environments. The full-size model has a stronger learning ability, which may depress the disturbance from scene information.

\vspace{-2mm}
\section{Conclusions}
\label{sec:conclusion}
The method of weight quantization is investigated on compressing DNN-based SV models. The experiments are carried out on two commonly used SV model structures: ECAPA and ResNet. The experiments show that the SV model size can be compressed by multiple times with a slight performance decline. The ResNet gives robust performance after quantization. The performance of ECAPA is affected more by low-bitwidth quantization than ResNet. The quantized models perform comparable or better than the full-precision models in a cross-language SV evaluation, indicating that the model’s generalization ability is preserved after model quantization. The quantized models retain speaker-relevant knowledge and obtain additional information with respect to the environmental sound. 





\bibliographystyle{IEEEtran}
\bibliography{mybib}

\end{document}